\documentclass{PoS}

\usepackage{dsfont}

\title{Density of states techniques for lattice field theories using the functional fit approach (FFA)}

\ShortTitle{Density of states techniques}

\author{
Christof Gattringer$^{\;a}\,$, Mario Giuliani$^{\;a}\,$, Alexander Lehmann$^{\;a,b}\,$, 
Pascal T\"orek$^{\;a}$ \thanks{$\;$ This work is based on the presentations by Mario Giuliani and Pascal T\"orek in the parallel sessions. 
It is supported by the Austrian 
Science Fund FWF, through the DK {\sl Hadrons in Vacuum, Nuclei, and Stars} (FWF DK W1203-N16) and by Grant I 1452-N27. Partial
support from DFG TR55, {\sl ``Hadron Properties from Lattice QCD''} is acknowledged. We thank Ydalia Delgado Mercado,
Kurt Langfeld, Biagio Lucini, 
Axel Maas, Michael M\"uller-Preussker and Uwe-Jens Wiese for discussions and comments on the literature.}\\ \\
$^a$Universit\"at Graz, Institut f\"ur Physik, Universit\"atsplatz 5, 8010 Graz, Austria \\
$^b$Humboldt-Universit\"at zu Berlin, Institut f\"ur Physik, 12489 Berlin, Germany \\ \\
        E-mail: \email{christof.gattringer@uni-graz.at}, \email{mario.giuliani@uni-graz.at}, \email{alehmann@physik.hu-berlin.de},
        \email{pascal.toerek@uni-graz.at}, }

\abstract{We discuss a variant of density of states (DoS) techniques for lattice field theories, the so-called "{\sl functional fit approach}'' (FFA). 
The DoS FFA is based on a density of states $\rho(x)$ which is parameterized on small intervals of the argument $x$ of  $\rho(x)$. On these intervals
restricted Monte Carlo simulations with an additional Boltzmann factor $\exp(\lambda x)$ allow to determine $\rho(x)$ very precisely by 
obtaining its parameters from fitting the Monte Carlo data to a known function of $\lambda$.  

\hskip7mm
We describe the method in detail and show its applicability in four different systems, three of which have a complex action problem: 
The SU(3) spin model with a chemical potential, U(1) lattice gauge theory, the $\mathds{Z}_3$ spin model with chemical potential, 
and 2-dimensional U(1) lattice gauge theory with a topological term. In all cases we compare to reference calculations, which partly were done 
in a dual formulation where the complex action problem is absent. In all four cases we find a very encouraging performance of the 
DoS FFA.}

\FullConference{The 33rd International Symposium on Lattice Field Theory\\
		14 -18 July 2015\\
		Kobe International Conference Center, Kobe, Japan}

\begin{document}

\section{Introduction}
It is well known that Monte Carlo simulations of lattice field theories at finite density or with a topological term are plagued by the complex
action problem: The action $S$ has an imaginary part and the Boltzmann factor $e^{-S}$ cannot be used as a probability weight 
in a stochastic process. A possible way out is the density of states approach, where a density $\rho(x)$ is computed in a simulation without
the imaginary part of the action which is then included in a subsequent reweighting step. Different variants of density of states techniques 
have seen quite some attention recently \cite{fodor}--\cite{FFA_Z3}. 

The main technical challenge of the density of states (DoS) approach is the determination of $\rho(x)$ with very high accuracy, which is necessary
because in the reweighting step for including the complex phase the density is probed with a highly oscillating factor. Achieving the 
required high accuracy is further complicated by the fact that $\rho(x)$ varies over many orders of magnitude and in some parameter 
ranges also the tails of $\rho(x)$ give a sizable contribution to physical observables. Consequently the recent developments have focussed
on achieving maximal accuracy for $\rho(x)$.

In this contribution we present and test an approach which we refer to as ''{\sl functional fit approach}'' (FFA). The argument $x$ of the density
is divided into small intervals and the density $\rho(x)$ is parameterized on each of these intervals. Subsequently one performs
a restricted Monte Carlo simulation on each of the intervals and probes the system with an additional Boltzmann factor
$e^{\, \lambda x}$, where the free parameter $\lambda$ is used to probe the system. The parameters of $\rho(x)$ 
on the given interval can be determined from fitting the restricted Monte Carlo data on the interval with a known function of $\lambda$.

\section{General scheme of the functional fit approach}

\subsection{Definition of the density}

\noindent
The basic equations for computing an observable $\langle O \rangle$ and the partition function $Z$ in a lattice field theory are given by
\begin{equation}
\langle O \rangle \; = \; \frac{1}{Z} \int \! D[\psi] \, e^{-S[\psi]} \, O[\psi] \quad , \qquad  Z \; = \; \int \! D[\psi] \, e^{-S[\psi]} \; .
\end{equation}
Here we use $\psi$ as a generic symbol for all fields of the theory (which for simplicity we assume to be bosonic here),
and $S[\psi]$ is the corresponding Euclidean lattice action. The observable is a functional $O[\psi]$ of the field configurations $\psi$. 
By $\int D[\psi]$ we denote the lattice path integral, i.e., the product over the measures of the fields at each site (or link) of the finite lattice. 

To keep the approach as general as possible we split the action into two parts as follows,
\begin{equation}
S[\psi] \; = \; S_\rho[\psi] \; + \; \xi \, X[\psi] \; .
\label{ssplit}
\end{equation}
Both parts $S_\rho[\psi]$ and  $X[\psi]$ are real functionals of the fields $\psi$ and $\xi$ is a parameter which is either real or purely imaginary,
where the latter case is the one relevant for a situation with a complex action problem. 
Here we use a weighted density of states and $S_\rho[\psi]$ is that part of the action which we include in the density. The second part 
$X[\psi]$ is then taken into account via reweighting, 
and in case of an imaginary $\xi$, this is how the complex action problem is handled. We remark that
even more general decompositions of the action are possible, where only part of the real part of the action is used in the weighted density 
and the rest of the real part is included via reweighting. 

Two examples for a decomposition according to (\ref{ssplit}) are:

\begin{itemize}

\item Charged scalar field $\phi$ with a chemical potential $\mu$:
\begin{equation}
S_\rho[\phi] \; = \; \mbox{Re} \; S[\phi] \quad , \qquad \xi X[\phi] \; = \; i \, \mbox{Im} \; S[\phi]  \quad , \qquad \xi \; = \; i \, \sinh(\mu) \; ,
\label{scalarmu}
\end{equation}
where $S_\rho$ and $X$ are essentially the real and imaginary parts of the action and the control parameter $\xi$ is given
by $\xi = i \, \sinh(\mu)$.

\item Yang-Mills theory with a topological term:  
\begin{equation}
S_\rho[U] \; = \; S_G [U] \quad , \qquad X[U] \; = \; Q[U] \quad , \qquad \xi \; = \; i\, \theta \; .
\label{ymtheta}
\end{equation}
We use $U$ to denote the configurations of the lattice gauge fields and $S_G [U]$ is the gauge action. 
$Q[U]$ is a discretization of the topological charge and $\theta$ the vacuum angle. 

\end{itemize}
The next step is to introduce a weighted density of states, defined as
\begin{equation}
\rho(x) \; = \; \int \! D[\psi] \, e^{-S_\rho[\psi]} \, \delta \left( X[\psi] - x \right) \; .
\end{equation} 
Obviously we obtain for the partition sum $Z$ and for observables $O$ which are functions $O[X]$:
\begin{equation}
Z \; = \; \int_{x_{min}}^{x_{max}} \!\! dx \; \rho(x) \, e^{- \, \xi \, x} \quad , \qquad \langle O \rangle \; = \;
\frac{1}{Z}  \int_{x_{min}}^{x_{max}} \!\! dx \; \rho(x) \, e^{- \, \xi \, x} \, O[x] \; ,
\end{equation}
where $x_{min}$ and $x_{max}$ denote the bounds of $X[\psi]$. Usually there is a symmetry transformation of the fields,
$\psi \, \rightarrow \, \psi^{\,\prime}$, with $S_\rho[\psi^{\,\prime}] = S_\rho[\psi]$, 
$X[\psi^{\,\prime}] = - X[\psi]$ and $\int D[\psi^{\,\prime}] = \int D[\psi]$. Such a symmetry 
guarantees the reality of $Z$ and implies that $\rho(x)$ is an even function,
\begin{equation}
\rho(-x) \; = \; \int \! D[\psi] \, e^{-S_\rho[\psi]} \, \delta \left(X[\psi] + x \right) \; = \; 
\int \! D[\psi^{\,\prime}] \, e^{-S_\rho[\psi^{\,\prime}]} \, \delta \left(- X[\psi^{\,\prime}] + x \right) \; = \; \rho(x) \; .
\end{equation}
It is easy to show that the expressions for $Z$ and $\langle O \rangle$ simplify to
\begin{equation}
Z \, = \, 2 \int_0^{x_{max}} \!\! dx \; \rho(x) \, \cosh( \xi \, x) \; , \; \langle O \rangle \, = \,
\frac{2}{Z}  \int_0^{x_{max}} \!\! dx \; \rho(x) \, \Big( \!\cosh( \xi \, x) \, O_e [x] - i  \sinh( \xi \, x) \, O_o [x] \!\Big)  \; ,
\label{evenobs}
\end{equation}
where we have defined the even and odd parts of the observable as $O_e[x] = (O[x] + O[-x])/2$, $O_o[x] = (O[x] - O[-x])/2$.
Note that for purely imaginary $\xi$ the sinh and cosh are replaced by sin and cos.
Eq.~(\ref{evenobs}) shows that we need $\rho(x)$ only for positive $x \in [0,x_{max}]$.

\subsection{Parametrization of the density}

\noindent 
The next step is to find a suitable parametrization of the density $\rho(x)$, which in many cases  
is an exponential ansatz which we discuss now. For the parametrization we first divide \cite{wanglandau} the range $[0,x_{max}]$
into $N$ intervals of variable size $\Delta_i, \, i = 0,1 \, ... \, N-1$ with $\sum_{i=0}^{N-1} \Delta_i = x_{max}$. 
For the boundaries of the 
intervals we introduce the notation $x_n = \sum_{i=0}^{n-1} \Delta_i $, such that the $n$-th interval is $[x_n, x_{n+1}]$. 
The exponential ansatz for $\rho(x)$ now has the form 
\begin{equation}
\rho(x) \; = \; e^{\, -l(x)} \qquad \mbox{with} \qquad  l(x) \; = \; d_n \, + \, x \, k_n \quad \mbox{for} \quad x \, \in \, [x_n, x_{n+1}] \; .
\label{rhoparam}
\end{equation}
In other words the density is the exponential of a piecewise linear function $l(x)$. For each interval this linear function is
parameterized by a constant $d_n$ and a slope $k_n$. We furthermore require $l(x)$ to be continuous and normalize it to $l(0) = 0$, which 
corresponds to the normalization $\rho(0) = 1$ (such a normalization can be chosen freely). The continuity condition and the normalization 
completely determine the constants $d_n$, and the slopes $k_n, \, n = 0,1 \, ... \, N-1$ are the only remaining parameters of $\rho(x)$.
A simple exercise gives the explicit representation 
\begin{equation}
l(x) \; = \; \sum_{i=0}^{n-1} \Delta_i \, ( k_i - k_n) \, + \, x \, k_n \quad \mbox{for} \quad x \, \in \, [x_n, x_{n+1}] \; .
\label{lfinal}
\end{equation}
We remark that one can use variable interval sizes $\Delta_i$ to achieve a finer parametrization of $\rho(x)$ in regions of $x$ where $\rho(x)$ 
shows a large variation. 

\subsection{Determination of the parameters with restricted expectation values}

\noindent 
For the determination of the parameters $k_n$ we define so-called restricted expectation values 
$\langle\!\langle O \rangle\!\rangle_n(\lambda)$. They depend on a real parameter $\lambda$ and are defined as \cite{langfeld}:
\begin{eqnarray}
\langle\!\langle O \rangle\!\rangle_n(\lambda) & = & \frac{1}{Z_n(\lambda)} \int \! D[\psi] \, e^{\, -S_\rho[\psi] \, + \, \lambda X[\psi]} \, 
O[X[\psi]] \, \Theta_n[X[\psi]] \; , 
\nonumber \\
Z_n(\lambda) & = & \int \! D[\psi] \, e^{\, -S_\rho[\psi] \, + \, \lambda X[\psi]} \, \Theta_n[X[\psi]] \; ,
\nonumber \\
\Theta_n[x] & = & \left\{ 
\begin{array}{cl}
1 & \; \mbox{for} \; x \, \in \, [x_n, x_{n+1}]  \\
0 & \quad \mbox{otherwise}
\end{array} \right. \; .
\label{restrictvev}
\end{eqnarray}
The function $\Theta_n[x]$ restricts the values of $x$ to the interval $[x_n, x_{n+1}]$ -- thus the name ''{\sl restricted expectation values}''. The 
parameter $\lambda$ enters the definitions of  $\langle\!\langle O \rangle\!\rangle_n(\lambda)$ and $Z_n(\lambda)$ via the additional 
Boltzmann factor $e^{\, \lambda X[\psi]}$ and allows to probe the system by varying $\lambda$. In the context of the density of states
approach this Boltzmann factor can be used to populate different regions of the density $\rho(x)$ in order to determine it accurately 
for all $x \in [x_n, x_{n+1}]$. The restricted expectation values $\langle\!\langle O \rangle\!\rangle_n(\lambda)$ can easily be evaluated in a Monte Carlo
simulation with an additional rejection step for configurations where $X[\psi] \notin [x_n, x_{n+1}]$.

One can now also work out $\langle\!\langle O \rangle\!\rangle_n(\lambda)$ and $Z_n(\lambda)$ using the density of states. One finds
\begin{equation}
\langle\!\langle O \rangle\!\rangle_n(\lambda)  \; = \; \frac{1}{Z_n(\lambda)} \int_{x_n}^{x_{n+1}} dx \, \rho(x) \, e^{\, \lambda \, x} \, O[x] 
\quad , \quad Z_n(\lambda) \; = \; \int_{x_n}^{x_{n+1}} dx \, \rho(x) \, e^{\, \lambda \, x} \; .
\label{onrho}
\end{equation}
Particularly simple is the expectation value $\langle\!\langle X \rangle\!\rangle_n(\lambda)$, which we write
as $\langle\!\langle X \rangle\!\rangle_n(\lambda) \; = \; \partial Z_n(\lambda) / \partial \lambda$.
Using the parametrization (\ref{rhoparam}), (\ref{lfinal}) of the density $\rho(x)$, we can explicitly evaluate 
$Z_n(\lambda)$ from (\ref{onrho}) and thus also $\langle\!\langle X \rangle\!\rangle_n(\lambda)$.
A straightforward calculation gives
\begin{equation}
\frac{1}{\Delta_n} \Big( \langle\!\langle X \rangle\!\rangle_n(\lambda) - x_n \Big) \, - \, \frac{1}{2} \;\; = \;\; F\big( (\lambda - k_n) \Delta_n \big) \; ,
\label{fitequation}
\end{equation}
where we have introduced the function 
\begin{equation}
F(s) \; = \; \frac{1}{1 - e^{-s}} \, - \, \frac{1}{s} \, - \, \frac{1}{2} \; .
\label{fitfunction}
\end{equation}
$F(s)$ is a real-valued smooth function of a real variable $s$. One finds $F(0) = 0$, $\lim_{s \rightarrow \pm \infty} F(s) = \pm \frac{1}{2}$ and
$F^{\, \prime}(0) = \frac{1}{12}$. The remarkable fact is that Eq.~(\ref{fitequation}) can be used to precisely determine the parameter $k_n$:
The left hand side of (\ref{fitequation}) can be evaluated with high accuracy in a restricted Monte Carlo simulation for several values of
$\lambda$. The result is then described by the right hand side of (\ref{fitequation}) where $k_n$ appears as the only undetermined 
variable, and we compute $k_n$ as the
result of a one-parameter fit of the data for the left hand side of (\ref{fitequation}) with the function $F\big( (\lambda - k_n) \Delta_n \big)$.
Thus the name ''{\sl functional fit approach}'' (FFA).

The procedure is then repeated for all intervals $[x_n, x_{n+1}]$ and we obtain all $k_n, \, n = 0,1 \, ... \, N-1$. From the $k_n$ we 
obtain the density $\rho(x)$ via (\ref{lfinal}) and (\ref{rhoparam}) and can compute the expectation values of observables 
using (\ref{evenobs}).

This concludes the general description of the FFA approach and we now discuss its actual application for several examples.

\section{The SU(3) spin system with a chemical potential}
\label{su3}

We begin the explicit presentation of the FFA approach by discussing its implementation for the SU(3) spin model with a chemical potential $\mu$. 
The model is an effective theory for the local Polyakov loop $P(x)$. It has a dual representation \cite{SU3a} and the dual simulation results
from \cite{SU3b} will be used as reference data for testing the FFA results. The action of the model is given by
\begin{equation}
S[P] \; = \; \tau \, \sum\limits_x \sum\limits_{\nu=1}^3 
\Bigl[ \mathrm{Tr}\, P(x) \, \mathrm{Tr}\, P(x+\hat\nu)^{\dagger} + c.c. \Bigr]
 \; + \; \kappa\,  \sum\limits_x \Bigl[ e^\mu \, \mathrm{Tr}\, P(x) + e^{-\mu} \, \mathrm{Tr}\, P(x)^{\dagger} \Bigr] \; .
 \label{su3action}
\end{equation}
Here $x$ runs over the sites of a 3-dimensional lattice with periodic boundary conditions. $\tau$ is a parameter that is an increasing function
of the temperature of the effective theory and $\kappa$ is a parameter that is a decreasing function of the quark mass.  The dynamical
degrees of freedom are the Polyakov loop variables $P(x) \in$ SU(3) attached to the sites $x$ of the lattice. Since they only enter with their 
trace, we can use the simplified representation 
$P(x) \, = \; \mbox{diag} ( e^{i \theta_1(x)}, e^{i \theta_2(x)},e^{-i(\theta_1(x)+\theta_2(x))})$ with only two angles 
$\theta_1(x), \theta_2(x) \in [-\pi, \pi]$. 

The decomposition of the action into $S_\rho[P]$ and $X[P]$ according to (\ref{ssplit}) 
is straightforward,  
\begin{equation}
S_\rho[P] \; = \; \mbox{Re} \, S[P] \; , \; \;
X[P] \; = \; \sum\limits_{x} \bigl[ \sin(\theta_1(x)) + \sin( \theta_2(x) ) - \sin(\theta_1(x) + \theta_2(x) ) \bigr] \; , \;  \;
\xi \; = \; i \, 2 \kappa \sinh(\mu) \; .
\end{equation}
Up to a factor the functional $X[P]$ is the imaginary part of the action. 
$X[P]$  is bounded, i.e., $X[P] \in [-\frac{3 \sqrt{3}}{2} V; \frac{3 \sqrt{3}}{2} V] \equiv [-x_{max},x_{max}]$, where $V$ is 
the number of lattice points. $X[P]$ is odd
under the transformation $\theta_i(x) \rightarrow - \theta_i(x)$, while $S_\rho[P]$ and the path integral measure $D[P]$ (the product 
over the reduced Haar measures) are even, such that (\ref{evenobs}) applies and we need the density $\rho(x)$ only for positive $x$.
The density of states is parameterized as in (\ref{rhoparam}) and (\ref{lfinal}), and
restricted vacuum expectation values are defined as given in (\ref{restrictvev}). The parameters $k_n$ of the density can be
determined from fits as outlined in (\ref{fitequation}), (\ref{fitfunction}). 

For the tests presented here we performed simulations on lattices of size $8^3$ at $\kappa = 0.005$ and several values 
of $\tau$ and $\mu$. The statistics used for computing each restricted expectation value in this exploratory study
is typically 5000 measurements separated by 10 updates 
for decorrelation after 2000 equilibration updates. We use between $N = 250$ and $N = 500$ intervals 
for dividing $[0,x_{max}]$ with typically two different sizes for the $\Delta_n$.

\begin{figure}[t]
\begin{center}
\hspace*{8mm}
\includegraphics[width=12cm]{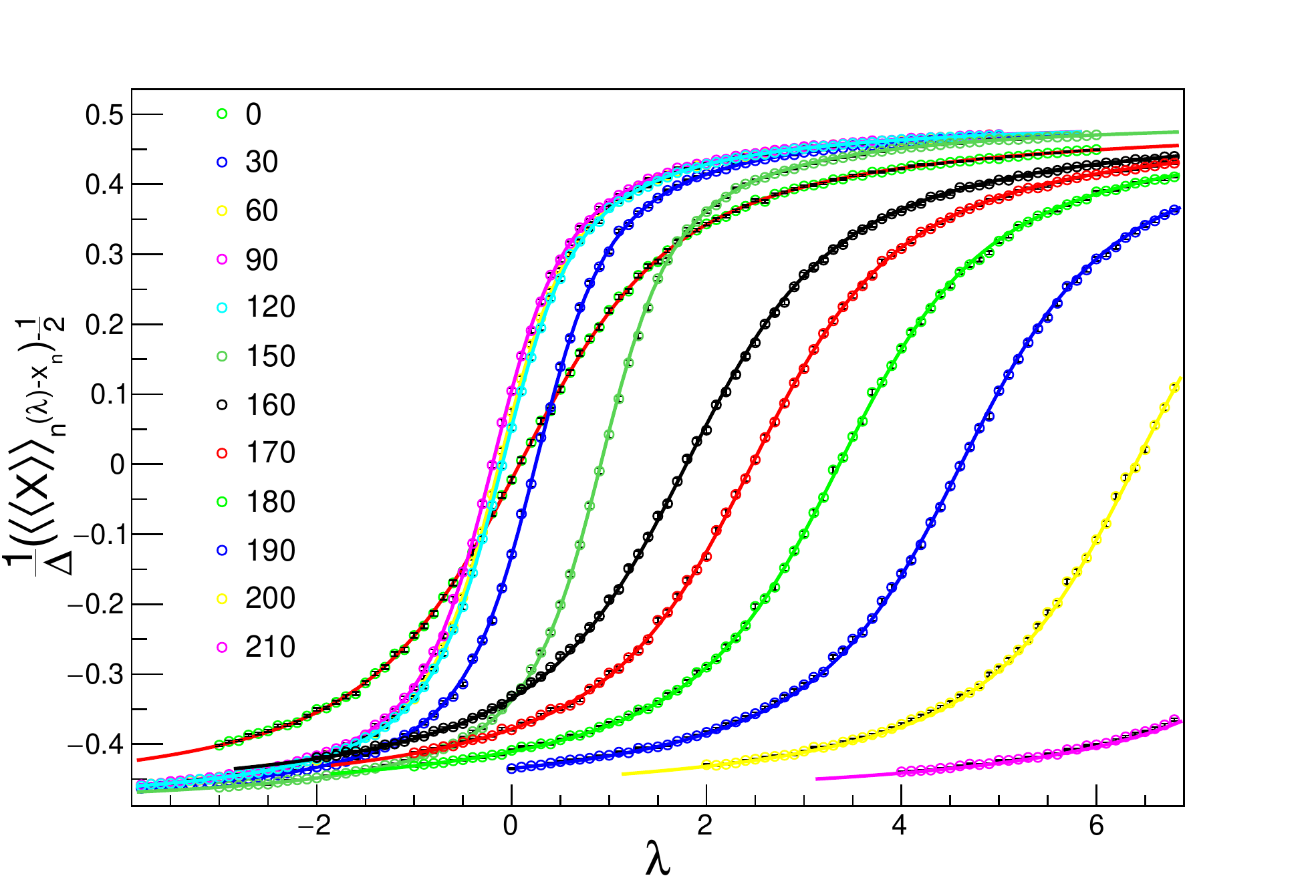} 
\end{center}
\vspace{-5mm}
\caption{Results for the restricted vacuum expectation values 
$( \langle\!\langle X \rangle\!\rangle_n(\lambda) - x_n )/\Delta_n - 1/2$ 
as a function of $\lambda$ for different intervals $n$ (indicated by the numbers in the legend). The symbols are the Monte Carlo results
and the full curves the fits with $F ( (\lambda - k_n) \Delta_n)$. We show results for $\tau = 0.075, \mu = 0.0$. Note that we use two different 
sizes $\Delta_n$ for the intervals, as can be seen from the fact that some of the data sets have a smaller slope near 0.}
\label{fig_SU3_vev}
\end{figure}

We begin the discussion of the numerical results with showing data for the restricted expectation values  
$( \langle\!\langle X \rangle\!\rangle_n(\lambda) - x_n)/\Delta_n \, - \, \frac{1}{2}$, i.e., the lhs.\ of Eq.~(\ref{fitfunction}). The corresponding 
Monte Carlo data, shown as circles in Fig.~\ref{fig_SU3_vev}, are for $\tau = 0.075, \mu = 0.0$, and we give the results for
many intervals $n$ as labelled in the legend. The smooth curves are the results for the corresponding fits with $F ( (\lambda - k_n) \Delta_n)$. 
It is obvious that the Monte Carlo data are very well described by these fits. In fact this is an important self-consistency check for 
the chosen parameterization of the density (compare (\ref{rhoparam})). We remark, that in Fig.~\ref{fig_SU3_vev} two different values 
were used for the intervals $\Delta_n$, which leads to a different slope of the curves near their zeros.

\begin{figure}[t]
\begin{center}
\includegraphics[width=9cm]{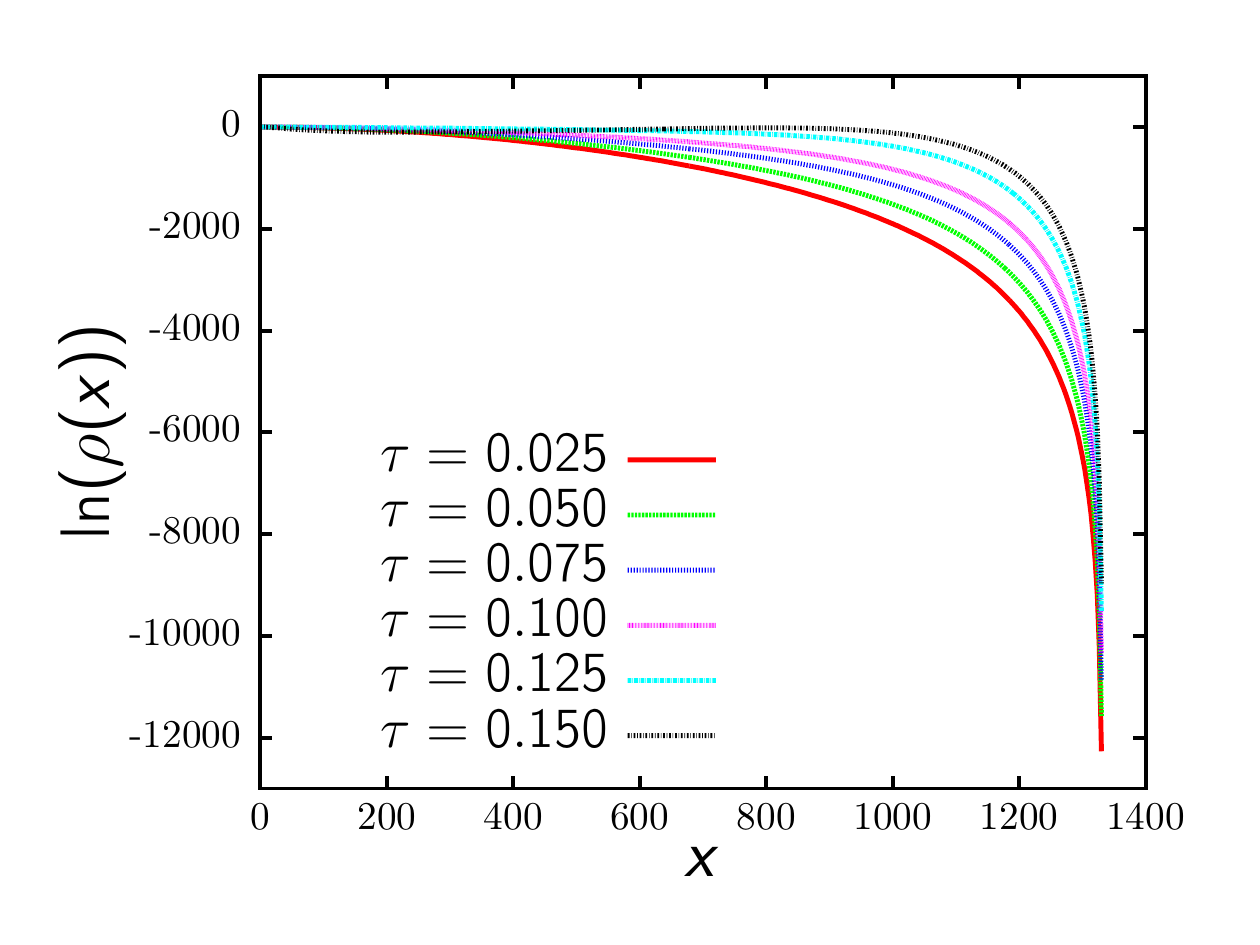} 
\end{center}
\vspace*{-10mm}
\caption{The logarithm of the density $\rho(x)$ versus $x$. We show results for $\mu = 0.0$ 
at several values of $\tau$.}
\label{fig_SU3_rho}
\end{figure}

From the fits to the restricted Monte Carlo data as shown in Fig.~\ref{fig_SU3_vev} we obtain all slopes $k_n$ and via (\ref{lfinal}) and 
(\ref{rhoparam}) the density $\rho(x)$. Our results for the density are shown in Fig.~\ref{fig_SU3_rho}, where we plot $\ln \rho(x)$ as a function
of $x$ for $\mu = 0.0$ and different values of the temperature parameter $\tau$. The plot shows that the density of states increases considerably
when increasing the temperature 
parameter for states with large $x$, i.e., states with large imaginary parts of the action. For all 
values of $\tau$ an obvious feature is that the density varies over many orders of magnitude, which illustrates the key challenge of all DoS
approaches. 

\begin{figure}[b]
\begin{center}
\hspace*{-9mm}
\includegraphics[height=6.2cm]{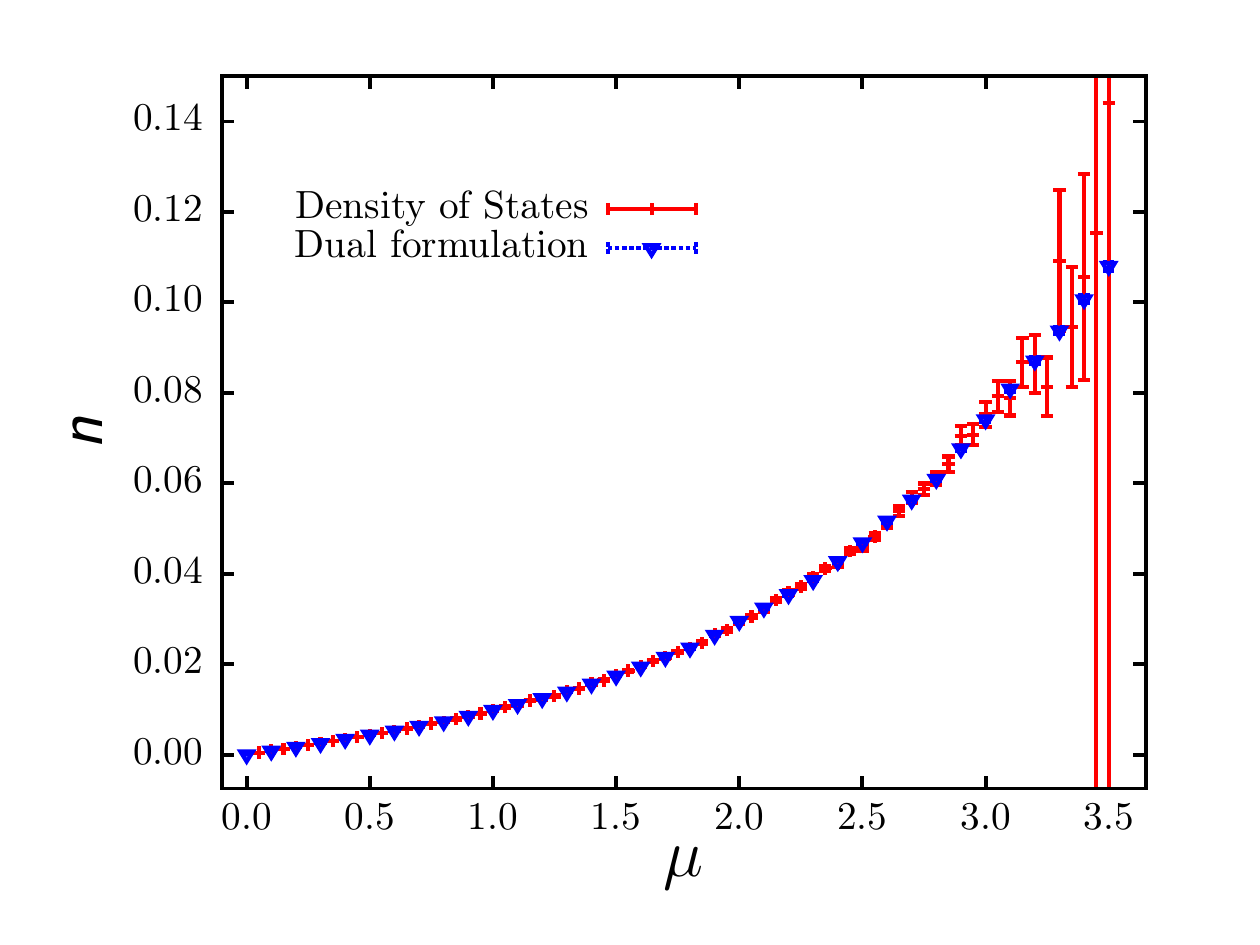}
\hspace{-8mm}
\includegraphics[height=6.2cm]{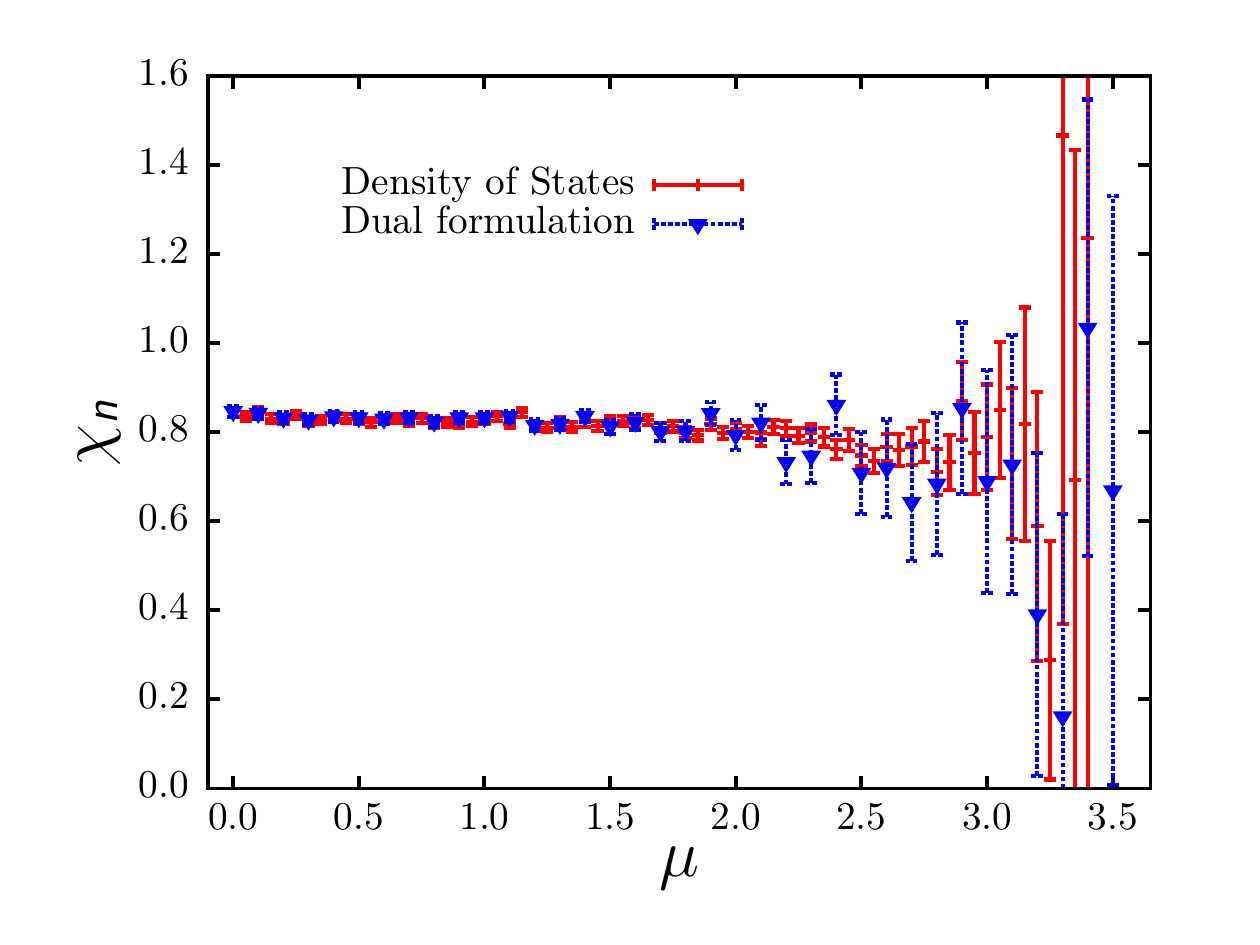}  
\end{center}
\vspace*{-8mm}
\caption{The particle number density $n$ (lhs.) and the particle number susceptibility $\chi_n$ (rhs.) versus the chemical potential $\mu$. 
We show the results for $\tau = 0.066, \kappa = 0.005$ lattices and compare the 
data from the FFA (plusses) to the results of a dual simulation (triangles).}
\label{fig_SU3_n}
\end{figure}

Having determined the density $\rho(x)$ the final step is the evaluation of observables according to (\ref{evenobs}). For the SU(3) spin model
we consider the particle number density $n$,
\begin{equation}
n \;\; \equiv \;\; - \frac{1}{V}\frac{1}{2\kappa} \frac{\partial}{\partial \sinh \mu} \, \ln Z \;\; = \;\; \frac{2}{V} 
\frac{1}{Z} \int_0^{x_{max}} \!\!\!  dx \; \rho(x) \, \sin( x \, 2\kappa \sinh \mu ) \, x \; ,
\end{equation}
and the corresponding susceptibility $\chi_n$ obtained by another derivative with respect to $\sinh \mu$. 
The results for these observables are shown in 
Fig.~\ref{fig_SU3_n} as a function of the chemical potential $\mu$. We find very good agreement of the DoS FFA results
with the results of the reference calculation with dual variables \cite{SU3b} up to chemical potential values of 
$\mu \sim 3$ for $n$ and up to $\mu \sim 2.5$ for $\chi_n$. We remark that the chemical potential $\mu$ as it appears in 
the action (\ref{su3action}) of the SU(3) model is rescaled by the inverse temperature $\beta$ of QCD, which is modeled 
by the effective SU(3) theory. In dimensionless units thus the range of good agreement between DoS FFA and dual methods 
is up to $\mu \beta \sim 3$ for the density and $\mu \beta \sim 2.5$ for  $\chi_n$, which is rather remarkable for the simple
exploratory implementation of FFA discussed here.

\section{The $\mathds{Z}_3$ spin system with a chemical potential}
\label{z3}

We now come to presenting the DoS FFA results for the $\mathds{Z}_3$ spin model in an external field which is closely related to 
the SU(3) spin system discussed in the previous section. The main difference is that the group SU(3) of the spins is replaced by its center
$\mathds{Z}_3$, such that the dynamical degrees of freedom are the spins $P_x\in \mathds{Z}_3 = \{1,e^{i2\pi /3},e^{-i2\pi /3}\}$, living on the sites
$x$ of a 3-dimensional lattice with periodic boundary conditions. The corresponding action is 
\begin{equation}
S[P]  =  \sum_x\left[\tau\sum_{\nu=1}^3 \big(P_x^\star P_{x+\hat{\nu}} + c.c. -2\big) + \kappa  e^{\mu\beta} (P_x-1)  + 
\kappa e^{-\mu \beta} (P_x^\star-1) \right] \; .
\label{action}
\end{equation}
The action is normalized such that $S[P]=0$ if $P_x=1\,\forall x$, and here we show explicitly the inverse temperature $\beta$ of the underlying
QCD-inspired theory, i.e., the chemical potential is coupled in the dimensionless form $\mu \beta$. The parameters $\tau$ and $\kappa$ play
the same role as in the SU(3) model.  Similar to the case of the SU(3) model also here one can decompose the action into real and
imaginary parts. The Boltzmann factor with the real part is used in the weighted density, while the phase from imaginary part needs to be integrated 
over the density for computing observables. 

\begin{figure}[t]
\centering
\hspace*{-3mm}
\includegraphics[height=50.mm]{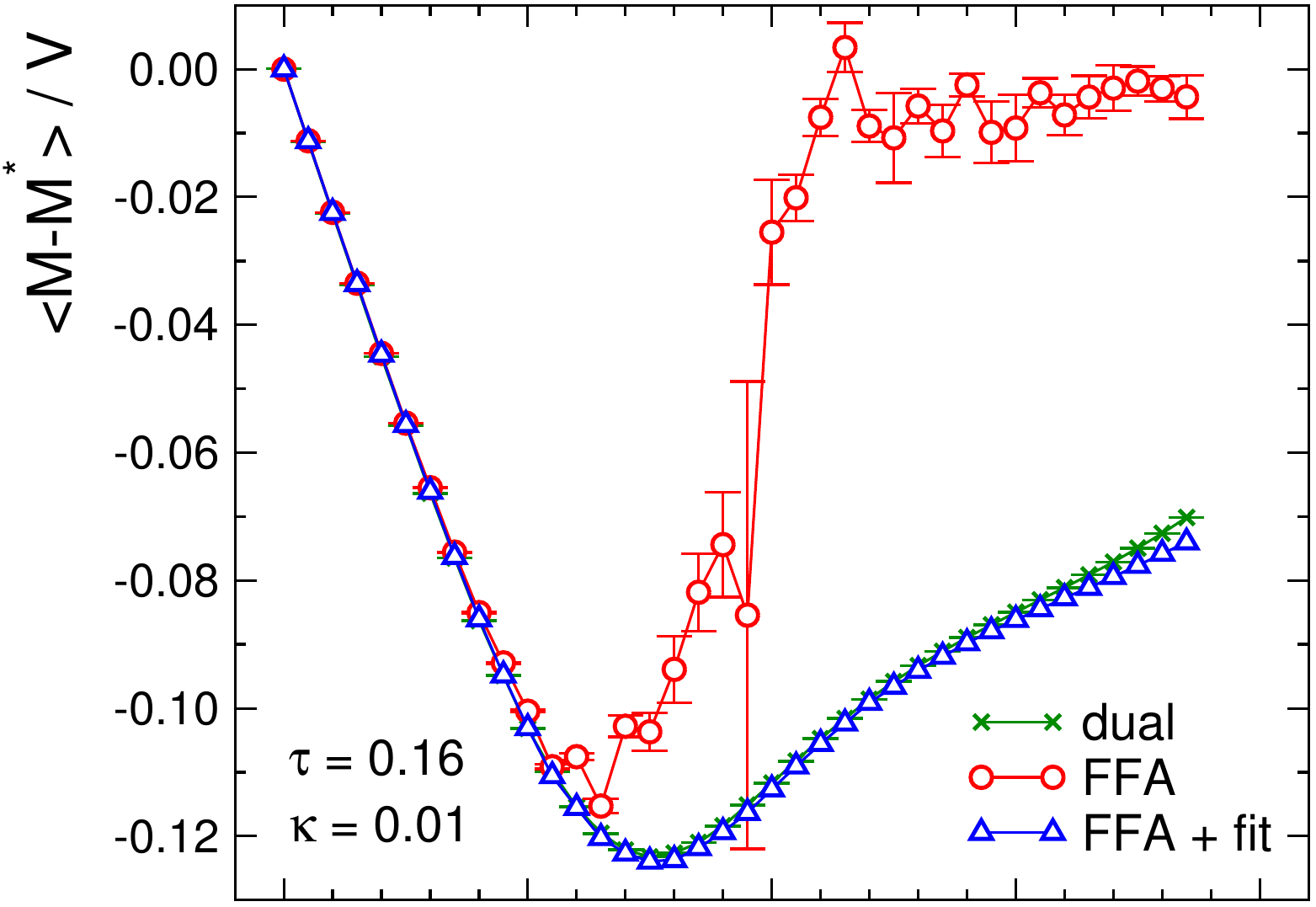}
\hspace{1mm}
\includegraphics[height=50mm]{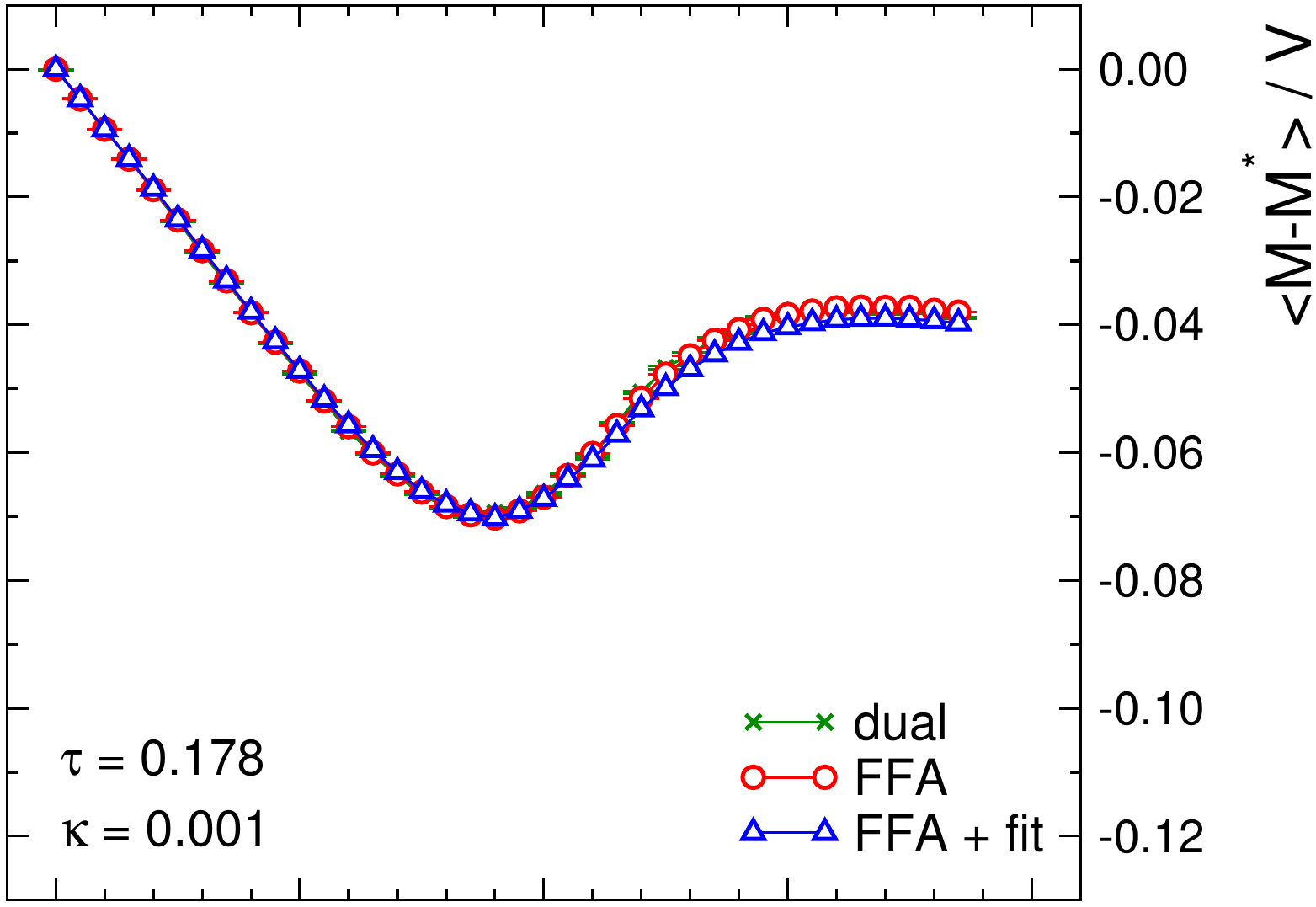}
\vspace{5mm}
\hspace*{-4mm}
\includegraphics[height=55.5mm]{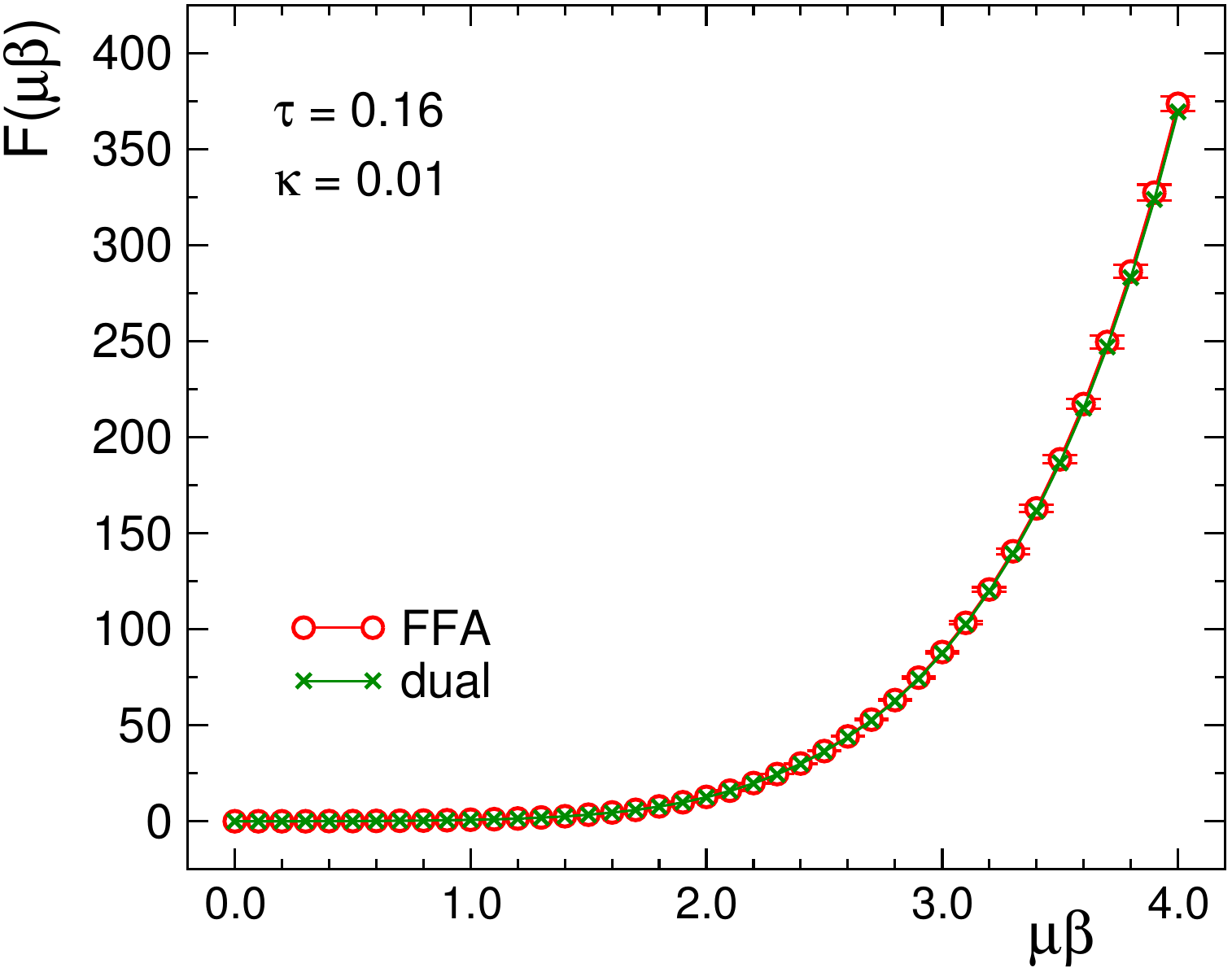}
\hspace{1mm}
\includegraphics[height=55.5mm]{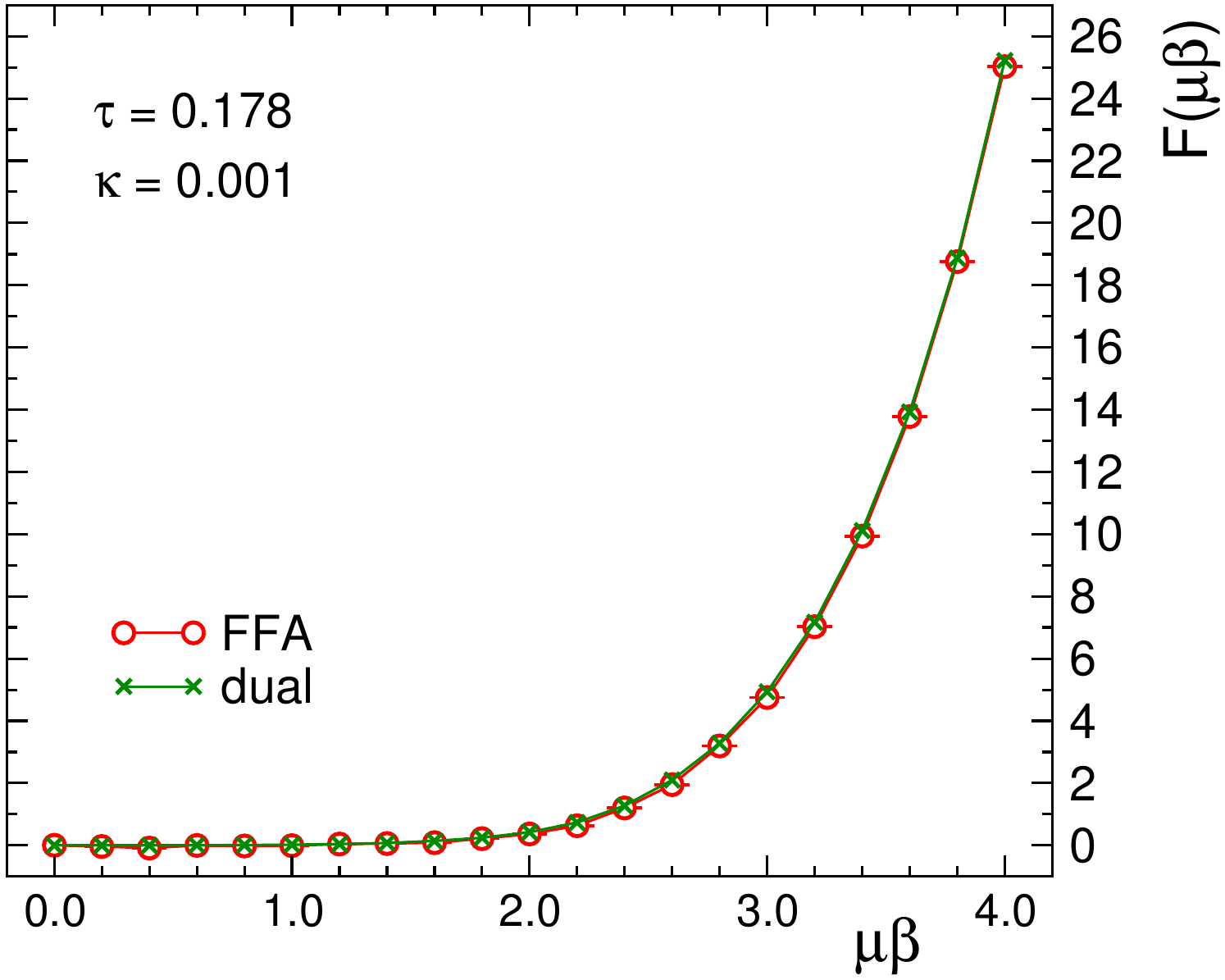}
\caption{Results for $\langle M-M^*\rangle$ (top row of plots) and the free energy (bottom)  
as a function of $\mu \beta$. We use  
two sets of parameters, $\tau = 0.16$, $\kappa = 0.01$ on the lhs., and $\tau = 0.178$ and $\kappa = 0.001$ on the rhs. 
We show results from the FFA algorithm, the FFA algorithm combined with a fit of $\rho(x)$ (only in the top row of plots) 
and for comparison also the 
results from a simulation in the dual representation.}
\label{observables}	
\end{figure}

There is an important difference to the SU(3) case discussed in the previous section: Here the dynamical degrees, i.e., 
the $\mathds{Z}_3$ spins, are discrete and the imaginary part which plays the role of the argument $x$ of the density $\rho(x)$ can have 
only a finite number of discrete values $x = -V, -V+1, \, .... \, V$ (on a finite volume $V$). 
Thus, parameterizing the density as piecewise constant, 
\begin{equation}
\rho(x) \; = \; \prod_{j=0}^{|x|} e^{-a_j} \; = \; \exp \bigg( -\sum_{j=0}^{|x|} a_j \bigg) 
\; , \; \; x = -V, -V+1, \, .... \, V -1, V \; ,
\label{rhoparamZ3}
\end{equation}
even allows for an exact representation of the density of states. For determining the parameters $a_j$ of the density we use a strategy \cite{FFA_Z3}
where we perform overlapping restricted Monte Carlo simulations which also extend to the two neighboring intervals next to the 
interval $j$ where we want to determine $a_j$. Also for this case one can work out the partition sum and the restricted expectation values
in terms of the parameterized density, and again the parameters $a_j$ can be obtained from fitting the expectation values 
with a known function of $\lambda$.

Here we present results \cite{FFA_Z3} for $\langle M - M^* \rangle$, where $M  =  \sum_x P_x$,  as well as for the free energy $F = - \ln Z$. 
In Fig.~\ref{observables} we show our results for two different values of the temperature parameter $\tau$, one in the confining phase of the theory,
the other one in the deconfined phase (see \cite{Dual_Z3}). In the lhs.\ column we plot the results
for $\tau = 0.16$ (confining), while the rhs.\ column is for $\tau = 0.178$ (deconfined). The top row displays $\langle M - M^* \rangle$, 
the bottom row is for the free energy. We compare the results from plain DoS FFA (circles) to the outcome of the dual simulation from
\cite{Dual_Z3} (crosses). For $\langle M - M^* \rangle$ we also show the results after the density was fit with a polynomial (triangles), 
a strategy which we discuss below.

It is obvious that for the free energy the DoS FFA and the dual results agree perfectly up to $\mu \beta = 4.0$. For $\langle M - M^* \rangle$
the agreement is reasonable only for the deconfined data set ($\tau = 0.178$, rhs.), while for the confined dataset 
($\tau = 0.16$, lhs.) the agreement of DoS FFA and dual results breaks down already at  $\mu \beta \sim 1$.
The reason for this breakdown are small fluctuations in the density coming from the statistical error of the Monte Carlo determination of the density.
As $\mu$ is increased, the density is probed by a faster oscillating factor and the fluctuations start to spoil the results (see the discussion in 
\cite{Dual_Z3}). 

To reduce the effect of the statistical fluctuations it has been proposed \cite{langfeld} to fit the density $\rho(x)$ with a polynomial in $x$, 
which can be chosen to be an even polynomial here, since $\rho(x)$ is even. We implemented this suggestion in our analysis of the results  
for $\langle M - M^* \rangle$ and indeed we find a drastic improvement of the agreement with the reference data from the dual simulation 
(see \cite{Dual_Z3} for the details). We stress at this point that the polynomial fit is not an essential ingredient of DoS FFA (or any other DoS variant), 
but simply an efficient method for reducing the effect of statistical errors. Alternatively one could also considerably increase the statistics
of the Monte Carlo simulation, however, at a much higher computational cost.

\section{U(1) lattice gauge theory}
\label{u1}

In this section we very briefly discuss the case of U(1) lattice gauge theory in 4 dimensions, which obviously is a theory without a 
complex action problem. Thus here the DoS FFA method is applied in a more conventional setting where the density is integrated 
with a non-oscillating Boltzmann factor. Nevertheless, also here the numerics is challenging since the Boltzmann factor and the density vary 
over many orders of magnitude.
The degrees of freedom are the gauge variables $U_\mu(x) \in$ U(1) on the links of a 4-d lattice with
periodic boundary conditions. For the action we use the usual Wilson plaquette action
\begin{equation}
S_G[U] \; = \; - \beta \sum_{x} \sum_{\mu < \nu} \, \mbox{Re} \, U_\mu(x) \, U_\nu (x + \hat \mu) \,
U_\mu(x + \hat \nu)^\star \, U_\nu(x)^\star \; .
\end{equation}

\begin{figure}[t]
\begin{center}
\hspace*{0mm}
\includegraphics[width=15cm]{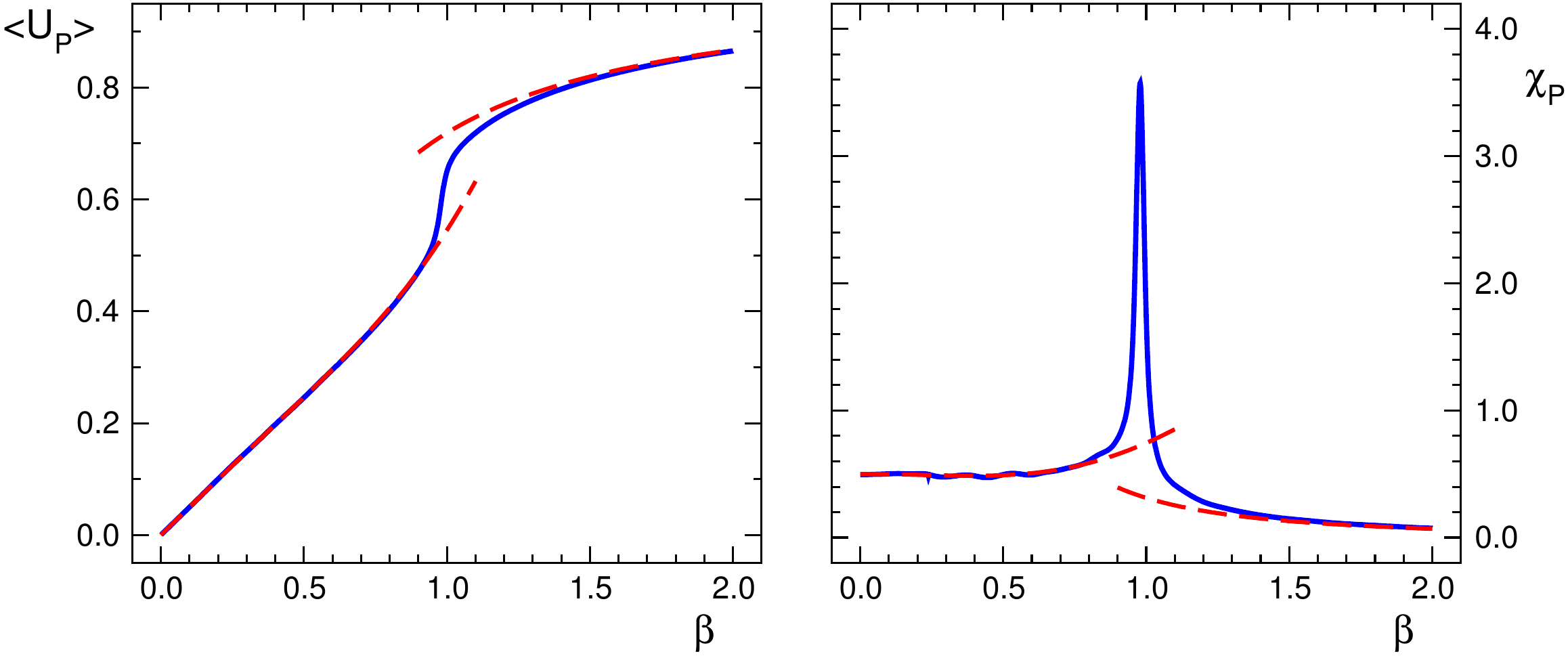} 
\end{center}
\vspace{-3mm}
\caption{Results for the the plaquette expectation value and the plaquette susceptibility as a function of $\beta$. We show
the results from DoS FFA (full curves) and from strong- and  weak coupling expansions (dashed) taken from 
\cite{lautrup}.}
\label{U1obs}
\end{figure}
The partition sum is given by $Z = \int D[U] e^{-S[U]}$, where $D[U]$ is the product over the U(1) Haar measures over all links of the lattice. 

For the implementation of DoS FFA according to the general scheme discussed in Section 2, we set $S_\rho = 0$, i.e., we use a density $\rho(x)$ 
without any weight factor and choose $\xi X[U] = S_G[U]$ with $\xi = \beta$. The parameterization of $\rho(x)$ is chosen as given 
in (\ref{rhoparam}) and (\ref{lfinal}). The determination of the parameters $k_n$ of $\rho(x)$ was implemented as described in Section 2. For 
a similar implementation of a DoS analysis of 4-dimensional U(1) lattice gauge theory with the DoS LLR approach see the last reference in 
$\cite{langfeld}$.

The observables we study here are the plaquette expectation value $\langle U_P \rangle$ and the corresponding susceptibility $\chi_P$ 
defined as
\begin{equation}
\langle U_P \rangle \; = \; \frac{1}{6V} \frac{\partial \ln Z}{\partial \beta} \quad , \qquad \chi_P \; = \; \frac{\partial \langle U_P \rangle}{\partial \beta} \; ,
\end{equation}
where $V$ denotes the number of lattice sites.

In Fig.~\ref{U1obs} we show our results for $\langle U_P \rangle$ and  $\chi_P$ as a function of $\beta$ for 
a small lattice ($V = 4^4$) and compare the DoS FFA results (full curves) to the  weak and strong coupling results from 
\cite{lautrup} (dashed curves). Our results show the well known (first order) transition near $\beta = 1.0$, and in the small- and large $\beta$ limits
we find good agreement with the strong- and weak coupling results.

\newpage
\section{U(1) lattice gauge theory in 2 dimensions with a topological term}
\label{u1topo}

Let us now finally come to an application of the DoS FFA to a class of systems where we expect a large potential for the DoS approach, 
i.e., models with a topological charge term as already outlined in the example given in Eq.~(\ref{ymtheta}). In particular the fact that the
vacuum angle $\theta$ is restricted to the range $\theta \in [0, \pi]$ makes the DoS approach interesting, since in the integral over the
density the oscillation frequency in the integrand $\rho(x) \cosh(\xi \, x)$ is linear in $\theta$ since $\xi = i \theta$, 
while for the chemical potential applications it is exponential in $\mu$ since $\xi = i \sinh \mu$ (compare Eq.~(\ref{scalarmu})).  

In this exploratory study we consider a system that is simpler than the Yang-Mills case sketched in (\ref{ymtheta}), namely the case of 
$U(1)$ lattice gauge theory in 2 dimensions with a topological term. On the lattice the gauge action $S_G[U]$ and the topological 
charge $Q[U]$ in the field theoretical definition can be written as (see \cite{topo2d} for our conventions for $Q[U]$):
\begin{equation}
S_G[U] \; = \; - \, \frac{\beta}{2} \sum_x \big[\, U_{x,p} \, + \, U_{x,p}^{\;*}\, \big]  \quad , \qquad
Q[U] \; = \; \frac{1}{i 4 \pi} \sum_x \big[ \; U_{x,p} \,  - \, U_{x,p}^{\; *} \, \big] \; ,
\label{u12daction}
\end{equation}
where the sums run over all sites $x$ of a 2-dimensional $N \times N = V$ lattice with periodic boundary conditions. 
$U_{x,p} \; = \; U_{x,1} \, U_{x+\hat{1},2} \, U_{x+\hat{2},1}^{\,*} \, U_{x,2}^{\,*}$ are the plaquettes made
from the link variables $U_{x,\nu} \in$ U(1). The continuum limit is obtained by $\beta \rightarrow \infty$, $V \rightarrow \infty$ 
at fixed ratio $\beta/V = const$, and we stress that only in this limit the partition sum 
$Z(\theta) = \int D[U] e^{-S_G[U] - i \theta Q[U]}$ becomes $2\pi$-periodic \cite{topo2d}. We remark that 
U(1) lattice gauge theory in 2 dimensions with a geometric definition of the topological charge $Q[U]$ was studied in \cite{wiese}, using a
Monte Carlo approach that includes a trial distribution for the topological charges and Metropolis updates connecting
neighboring charge sectors.
 
The transcription of the system to the general DoS FFA scheme as outlined in Section 2 is obvious via the identifications
$S_\rho[U] = S_G[U]$, $X[U] = Q[U]$ and $\xi = i \theta$. The parametrization we use is given by  
(\ref{rhoparam}) and (\ref{lfinal}) and implementing the FFA scheme for determining the $k_n$ as outlined in Section 2.3 is straightforward.

An important aspect for the choice of the model (\ref{u12daction}) as a test case for systems with a vacuum angle is the fact that, 
based on a dual variable approach, one can write the partition sum $Z$ as \cite{topo2d}
\begin{equation}
Z(\theta) \; = \; \sum_{q=-\infty}^{+\infty} \left[ \; {I}_{|q|}\big(2\sqrt{\eta \overline{\eta}}\big) 
\left(\sqrt{\frac{\eta}{\overline{\eta}}}\,\right)^{\!\!q} \; \right]^{V} \; ,
\label{Zgauge}
\end{equation}
where the $I_n$ are the modified Bessel functions and we define
$\eta\equiv\frac{\beta}{2}-\frac{\theta}{4\pi}$,  $\bar{\eta}\equiv\frac{\beta}{2}+\frac{\theta}{4\pi}$. Explicit expressions for the topological 
charge density and the topological susceptibility are then obtained as derivatives 
\begin{equation}
\langle q \rangle \; \equiv \;  \frac{-1}{V} \frac{\partial}{\partial \theta} \ln Z \quad , \qquad \chi_{t} 
\; \equiv \; \frac{-1}{V} \frac{\partial^2}{\partial \theta^2} \ln Z \; ,
\label{topobs}
\end{equation}
while for the representation in terms of the density of states we again use (\ref{evenobs}). However, the explicit expression (\ref{Zgauge}) 
even allows for an expression for the density $\rho(x)$ as a Fourier transform,
\begin{equation}
\rho(x) \; = \; \int d \theta  \, e^{i \theta x}  \, Z(\theta)  \; = \;  
\sum_{q=-\infty}^{+\infty} \int d \theta \, e^{i \theta x}  \left[ \; {I}_{|q|}\big(2\sqrt{\eta \overline{\eta}}\big) 
\left(\sqrt{\frac{\eta}{\overline{\eta}}}\,\right)^{\!\!q} \; \right]^{V} \; .
\label{rhoexplicit}
\end{equation}
The expression (\ref{rhoexplicit}) can be evaluated numerically with Mathematica and in this presentation we focus on the density 
$\rho(x)$ for assessing the quality of the results from DoS FFA. 

\begin{figure}[t]
\begin{center}
\hspace*{-8mm}
\includegraphics[height=6.12cm]{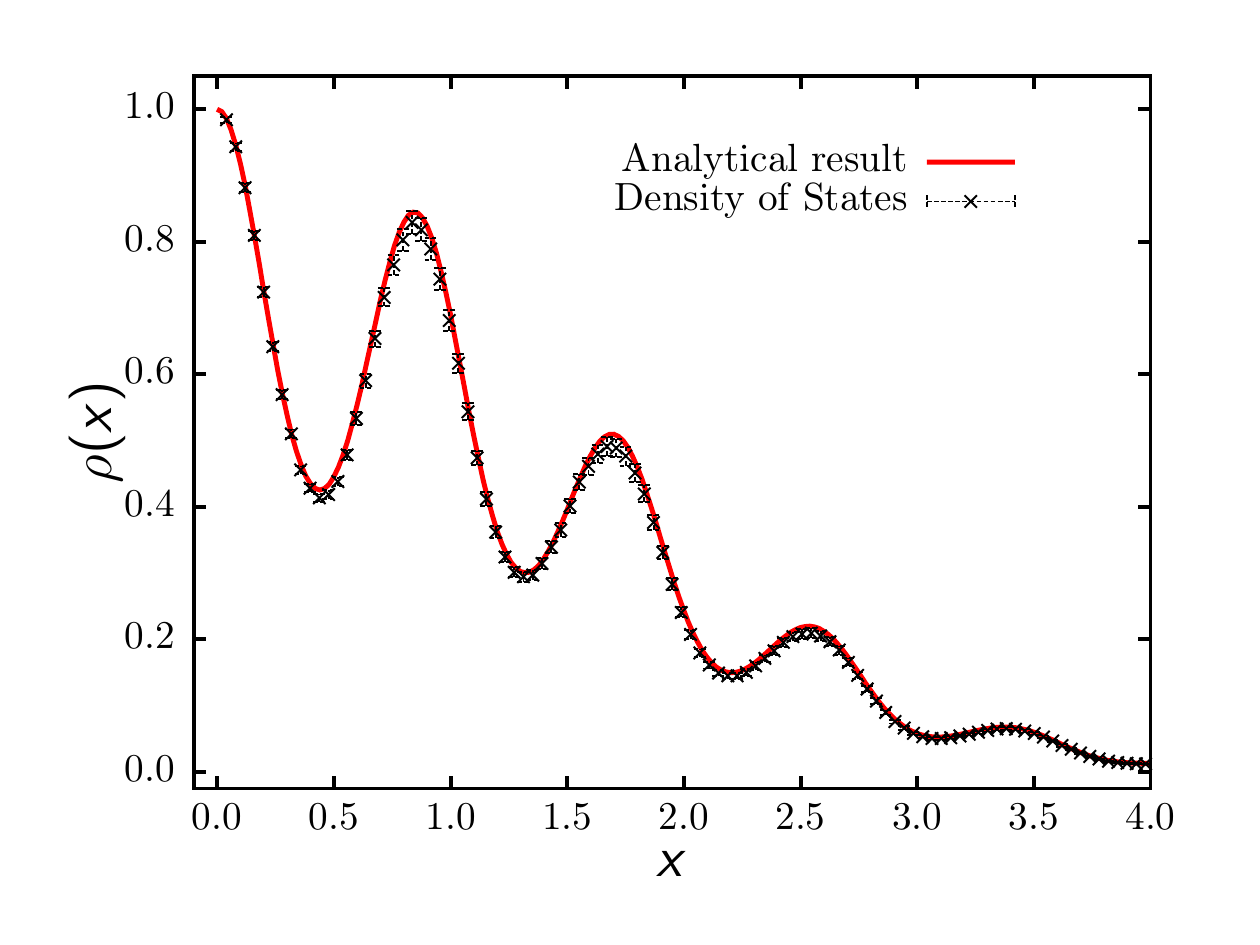}
\hspace{-7mm}
\includegraphics[height=6.12cm]{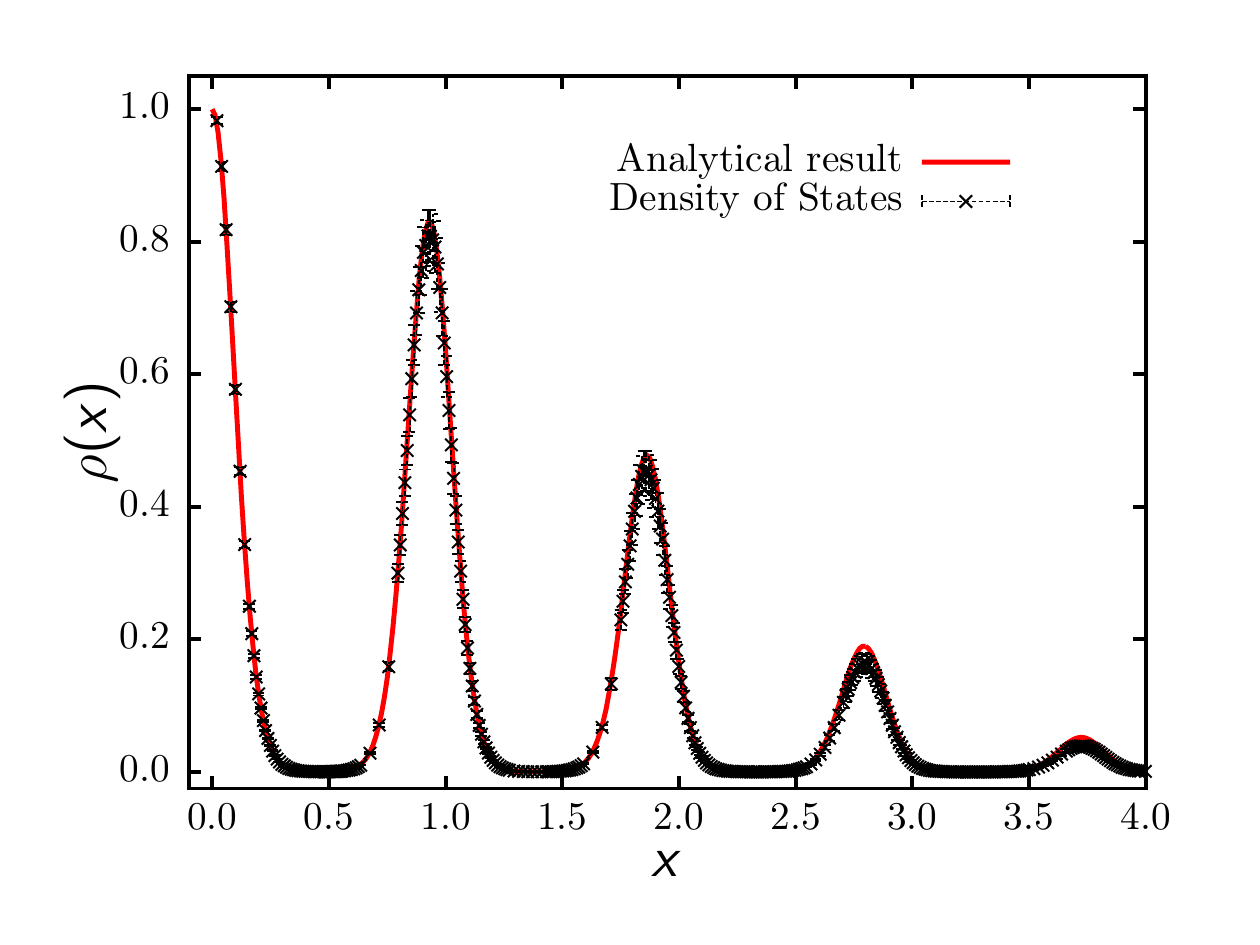}  
\end{center}
\vspace*{-8mm}
\caption{The density $\rho(x)$ as a function of $x$ for two different volumes and values of $\beta$: $20 \times 20, \, \beta = 4.0$ in the lhs.\ plot
and $28 \times 28, \, \beta = 7.84$ on the rhs. The crosses represent the results
from DoS FFA and the full curve the exact result from Eq.~(6.4).}
\label{fig_rho_U(1)}
\end{figure}

In this proceedings contribution we present only preliminary data from an exploratory study of the DoS FFA for U(1) lattice gauge theory with 
topological term. More specifically we use two lattice volumes $20 \times 20$ and $28 \times 28$ at $\beta = 4.0$ and $\beta = 7.84$, 
such that $\beta/V = 0.01 = const$. We currently work on results for bigger lattices and larger $\beta$, 
such that the system is closer to the continuum limit. In Fig.~\ref{fig_rho_U(1)} we show the DoS FFA results (crosses) for the 
density $\rho(x)$ as a function of $x$. The continuous curves are the reference results from (\ref{rhoexplicit}). 
 
It is obvious that the DoS FFA data agree very well with the reference curves from (\ref{rhoexplicit}). Only near the extrema of $\rho(x)$ we observe 
small deviations. These are due to the fact that there the slopes $k_n$ used in the parameterization (\ref{rhoparam})) are changing rapidly
near the extrema. Thus the observed discrepancy is a hint that near the extrema smaller interval sizes $\Delta_n$ should be used -- 
an improvement we are currently testing.

We have already remarked that the continuum limit is approached by sending $V$ and $\beta$ to infinity at a fixed ratio $\beta/V = const$, and only
in this limit the observables become $2\pi$-periodic. As a consequence one expects that $\rho(x)$ becomes sharply
peaked at integer values of $x$. The two datasets shown in Fig.~\ref{fig_rho_U(1)} both correspond to $\beta/V = 0.01$, with the rhs.~data at 
$V = 28 \times 28, \beta = 7.84$ being closer to the continuum limit than the lhs.~with $V = 20 \times 20, \beta = 4.00$. 
It is obvious that for the rhs.~plot the density is
indeed more peaked and in between the peaks $\rho(x)$ is approaching 0. Still we find a good description of the density from (\ref{rhoexplicit})
by the data from DoS FFA, despite the fact that here we can present only very preliminary data.

We stress at this point that regions with vanishing $\rho(x)$ do not pose a 
problem for the DoS FFA. The actual simulation is done with the ensemble (\ref{restrictvev}) which is modified by the additional Boltzmann factor
$e^{\lambda X}$ and the parameters of the density are determined from fitting the expectation values (\ref{fitfunction}) 
(compare also Fig.~\ref{fig_SU3_vev}). The fact that $\rho(x)$ may vanish for some $x$ has no influence on that procedure. We are currently
improving the technology for the application of DoS FFA to theories with a topological term and in particular study the performance of the 
DoS FFA in the 2-dimensional U(1) theory with vacuum term 
closer to the continuum limit. 

\newpage
\section{Concluding remarks}

In this proceedings contribution we give an overview over recent developments for the DoS FFA method and its application in various
systems. The DoS FFA method is characterized by a parameterization of the density of states $\rho(x)$ on small intervals of its argument $x$.
Restricted Monte Carlo simulations are performed on these intervals of size $\Delta_n$ 
and the system is probed with an additional Boltzmann factor $e^{\lambda x}$.
Varying $\lambda$ explores all possible states in the given interval of $x$ and the parameters used in the parameterization of $\rho(x)$ in 
that interval can be determined by fitting the Monte Carlo data with a known function of $\lambda$.  We stress that the method makes use of 
all Monte Carlo data that were created for the different values of $\lambda$, as they all contribute to the fit where the parameters of $\rho(x)$ 
are determined. Furthermore, one can also analyze the quality of that fit. If the quality is poor this is a signal that the chosen parameterization 
is too coarse and the $\Delta_n$  should be smaller. Thus the DoS FFA comes with a consistency check provided by the quality of the fit
of the restricted Monte Carlo data with the known function of $\lambda$.

The method was tested in four different systems with different types of complex action problems and different technical challenges. The first
system we explore is the SU(3) spin system with a chemical potential. The real part of the action is included in the weighted density of states 
which is determined in a straightforward way with the DoS FFA strategy. We show that the restricted expectation values are well described by 
the fit function -- and the quality of the fit indicates that the density is properly described by the chosen parameterization 
and interval size. We present results for the particle number and its susceptibility and find good agreement with the reference data 
from a dual simulation for up to at least $\mu \beta \sim 2.5$ -- quite remarkable for a first test with low statistics. 

The next system we analyzed is the $Z_3$ system where the degrees of freedom are discrete spins. Thus the imaginary part of the action
can have only discrete values and an exact parameterization can be chosen for $\rho(x)$ in
the form of a piecewise constant ansatz. Here one uses overlapping restricted simulations on neighboring intervals and determines the constants
again by fitting a known function of $\lambda$. For the free energy we here find perfect agreement with the dual reference simulation up to 
$\mu \beta \sim 4.0$, while for the imaginary part of the magnetization (related to the particle number) we find that for our $\tau = 0.16$ ensemble 
a fit of the density with a polynomial was needed to achieve the same accuracy.

A small test simulation was implemented for pure U(1) gauge theory in 4 dimensions which has no complex problem such that 
a density without weights was used and the numerical challenge comes from accurately integrating the density over many orders of magnitude.
Also in this case of a simple gauge theory we find good performance of the DoS FFA method. 

Finally we applied the DoS FFA method to a model system where the complex action problem is due to a topological term: U(1) gauge theory 
in two dimensions with a topological term. We believe that this type of systems are interesting for DoS techniques, since the frequency 
of the oscillating factor grows linearly with the vacuum angle $\theta$, while for systems with a chemical potential $\mu$ the frequency
grows exponentially in $\mu$. However, here other challenges have to be dealt with, in particular a density $\rho(x)$ which is non-monotonical and 
develops Dirac delta like behavior in the continuum limit. Thus one needs a very fine parameterization of $\rho(x)$ near the peaks of the
density. In the exploratory feasibility study presented here we show that these challenges can successfully be dealt with and we find
good agreement of the density from DoS FFA with the analytical result.

\end{document}